\documentclass[11pt]{article}
\usepackage{graphicx}
\usepackage{amsfonts}
\usepackage{theorem}
\newtheorem{Lem}{Lemma}[section]
\newtheorem{Def}[Lem]{Definition}
\newtheorem{The}[Lem]{Theorem}
\newtheorem{Prop}[Lem]{Proposition}

\newtheorem{Ax}[Lem]{Axiom}
\newcommand{\qed}{\hbox{\rule{6pt}{6pt}}}

\begin{document}

% paper title
\title{An axiomatic characterization of a two-parameter extended relative entropy}

\author{Shigeru Furuichi$^1$\footnote{E-mail:furuichi@chs.nihon-u.ac.jp}\\
$^1${\small Department of Computer Science and System Analysis,}\\
{\small College of Humanities and Sciences, Nihon University,}\\
{\small 3-25-40, Sakurajyousui, Setagaya-ku, Tokyo, 156-8550, Japan}}
\date{}
\maketitle
{\bf Abstract.} The uniqueness theorem for a two-parameter extended relative entropy is proven.
This result extends our previous one, the uniqueness theorem for a one-parameter extended relative entropy,
to a two-parameter case.
In addition, the properties of a two-parameter extended relative entropy are studied.
\vspace{3mm}

\vspace{3mm}


\vspace{3mm}

%%%%%%%%%%%%%%%%%%%%%%%%%%%%%%%%%%%%%%%%%%%%%%%%%%%%%%%%%%%%%%%%%%%%%%%%%%%%%%%%%%%%%%%%%%%%%%%%%%%%%%%%%%%%%%%%%%%%%%%%%%%%%%%%%%%%%%%%%%%%%%%%%%%%%%%%%%%%%%%
%%%%%%%%%%%%%%%%%%%%%%%%%%%%%%%%%%%%%%%%%%%%%%%%%%%%%%%%%%%%%%%%%%%%%%%%%%%%%%%%%%%%%%%%%%%%%%%%%%%%%%%%%%%%%%%%%%%%%%%%%%%%%%%%%%%%%%%%%%%%%%%%%%%%%%%%%%%%%%%
%%%%%%%%%%%%%%%%%%%%%%%%%%%%%%%%%%%%%%%%%%%%%%%%%%%%%%%%%%%%%%%%%%%%%%%%%%%%%%%%%%%%%%%%%%%%%%%%%%%%%%%%%%%%%%%%%%%%%%%%%%%%%%%%%%%%%%%%%%%%%%%%%%%%%%%%%%%%%%%
%%%%%%%%%%%%%%%%%%%%%%%%%%%%%%%%%%%%%%%%%%%  Section1  %%%%%%%%%%%%%%%%%%%%%%%%%%%%%%%%%%%%%%%%%%%%%%%%%%%%%%%%%%%%%%%%%%%%%%%%%%%%%%%%%%%%%%%%%%%%%%%%%%%%%%%%
%%%%%%%%%%%%%%%%%%%%%%%%%%%%%%%%%%%%%%%%%%%%%%%%%%%%%%%%%%%%%%%%%%%%%%%%%%%%%%%%%%%%%%%%%%%%%%%%%%%%%%%%%%%%%%%%%%%%%%%%%%%%%%%%%%%%%%%%%%%%%%%%%%%%%%%%%%%%%%%
%%%%%%%%%%%%%%%%%%%%%%%%%%%%%%%%%%%%%%%%%%%%%%%%%%%%%%%%%%%%%%%%%%%%%%%%%%%%%%%%%%%%%%%%%%%%%%%%%%%%%%%%%%%%%%%%%%%%%%%%%%%%%%%%%%%%%%%%%%%%%%%%%%%%%%%%%%%%%%%
%%%%%%%%%%%%%%%%%%%%%%%%%%%%%%%%%%%%%%%%%%%%%%%%%%%%%%%%%%%%%%%%%%%%%%%%%%%%%%%%%%%%%%%%%%%%%%%%%%%%%%%%%%%%%%%%%%%%%%%%%%%%%%%%%%%%%%%%%%%%%%%%%%%%%%%%%%%%%%%
\section{Introduction}
Shannon entropy \cite{Sha} is one of fundamental quantities in classical information theory 
and uniquely determined by the 
Shannon-Khinchin axiom or the Faddeev axiom. 
One-parameter extensions for Shannon entropy have been studied by many researchers \cite{AD}.
The R\'enyi entropy \cite{Ren} and the Tsallis entropy \cite{Tsa} are famous.
In the paper \cite{Suy}, the uniqueness theorem for the Tsallis entropy was proven.
See also the paper \cite{Csi} and the references therein, for the axiomatic characterizations of one-parameter extended entropies.
The two-parameter family of entropy was introduced by Borges and Roditi in \cite{BR} by the use of
the generalized Jackson derivative method.
Recently, a two-parameter extended entropy, which is essentially same form with the two-parameter family of entropy in \cite{BR},  has studied by 
several researchers \cite{KLS1,KLS2,WS} and 
the uniqueness theorem for a two-parameter extended entropy was proven in \cite{WS} by generalizing
the Shannon-Khinchin axiom. 
In this paper, we denote a two-parameter extended entropy by
$$
S_{\alpha,\beta}(x_1,x_2,\cdots,x_n)= \sum_{j=1}^n \frac{x_j^{\alpha}-x_j^{\beta}}{\beta-\alpha},
\quad (\alpha \neq \beta)
$$
for two real numbers $\alpha$ and $\beta$ such that 
$0\leq \alpha \leq 1 \leq \beta$ or $0 \leq \beta \leq 1 \leq \alpha$.
If we take $\alpha = 1$ or $\beta =1$, then it recovers
the Tsallis entropy defined by 
$$
S_{q}(x_1,x_2,\cdots,x_n)\equiv \sum_{j=1}^n\frac{x_j-x_j^q}{q-1},\quad (1\neq q \geq 0).
$$
The Tsallis entropy recovers Shannon entropy
$$
S_1(X)\equiv -\sum_{j=1}^n x_j \log x_j 
$$
 in the limit $q \to 1$.

In this paper, we study on information measure (entropy) defined for two probability distributions.
The relative entropy (Kullback-Leibler information or divergence) is defined for {\it two} probability distributions
 $X=\left\{x_1,\cdots,x_n\right\}$ and $Y=\left\{y_1,\cdots,y_n\right\}$:
$$D_1(X\vert \vert Y)\equiv \sum_{j=1}^n x_j \left(\log x_j -\log y_j\right).$$
Since Shannon entropy is defined for {\it one} probability distribution
and it can be reproduced by the relative entropy as
$\log n -D_1(X\vert \vert U) $ for the uniform distribution $U=\left\{1/n,\cdots,1/n\right\}$,
 the relative entropy can be regarded as a generalization for Shannon entropy.
We here note that we have one-parameter extended relative entropies 
such as the R\'enyi relative entropy $D^{R}_q(X\vert Y)$, 
$\alpha$-divergence $D^{(\alpha)}(X\vert \vert Y)$ and the Tsallis entropy $D^{T}_q(X\vert \vert Y)$.
These are defined by
\begin{eqnarray*}
&&D^{R}_q(X \vert \vert  Y) \equiv \frac{1}{q-1}\log \sum_{j=1}^n x_j^q y_j^{1-q}, \\
&&D^{(\alpha)}(X\vert \vert Y) \equiv \frac{4}{1-\alpha^2}\left(1-\sum_{j=1}^n x_j^{\frac{1-\alpha}{2}}y_j^{\frac{1+\alpha}{2}}\right), \\
&&D^{T}_q(X\vert \vert Y) \equiv\sum_{j=1}^n  \frac{x_j- x_j^qy_j^{1-q}}{1-q}, 
\end{eqnarray*}
for $q \neq 1$ and $\alpha \neq \pm 1$.
These quantities recover the relative entropy in their limit $q \to 1$ or $\alpha \to \pm 1$.
These quantities are also essentially same one in the sense that
\begin{eqnarray*}
&&D^{(q)}(X\vert \vert Y) = \frac{1}{q}D_q^T(X\vert \vert Y),  \quad (q \neq 0,1 ),\\
&&D_q^R(X\vert \vert Y) = \frac{\log\left\{1+(q-1)D_q^T(X\vert \vert Y)\right\}}{q-1},\quad (q \neq 1), 
\end{eqnarray*}
where we set $q=\frac{1-\alpha}{2}$ in $D^{(q)}(X\vert \vert Y)$. Here, we note that
the form $\sum_{j=1}^n x_j^qy_j^{1-q}$ is appeared in all one-parameter extended relative entropies.
Therefore it was sufficient to study one quantity of them, for the study of a one-parameter extension of the relative entropy.
It is also notable that the Tsallis entropy can be rewritten by the Tsallis relative entropy as a special case:
$$
S_q(X) = \ln_q n -n^{1-q} D_{q}(X\vert\vert U)
$$
for the uniform distribution $U=\left\{1/n,\cdots,1/n\right\}$, where the $q$-logarithmic function is defined by 
$$
\ln_q(x) \equiv \frac{x^{1-q}-1}{1-q},\,\, q>0,q\ne 1,x>0.
$$
Thus the uniqueness theorem for the Tsallis relative entropy was proven in our previous paper \cite{Furu1}. 

In the present paper, as a further extension of our previous result, 
we give a two-parameter extended axiom for the function defined for any pairs of the probability distributions 
and prove the uniqueness theorem for a two-parameter extended relative entropy.
This paper is organized as follows.
In Section 2, we review the uniqueness theorem for relative entropy proven by A.Hobson, and 
the uniqueness theorem for a one-parameter extended relative entropy.
In Section 3, we show our main theorem. In Section 4, we characterize the constant appeared in Section 3.
In Section 5, we give properties for a two-parameter extended relative entropy.

\section{Review of the uniqueness theorem for one-parameter extended relative entropy}

The uniqueness theorem for relative entropy was shown by A. Hobson as follows \cite{Hob}:
\begin{The}{\bf (\cite{Hob})}\label{axiom_Hob}
The function $D_1(A\vert \vert B)$ is assumed to be defined for any two probability distributions $A=\left\{a_j\right\}$ and $B=\left\{b_j\right\}$ for $j=1,\cdots ,n$.
If $D_1(A \vert\vert  B)$ satisfies the following conditions (R1)-(R5), then it is given by the form $k \sum_{j=1}^na_j\log\frac{a_j}{b_j}$ with a positive constant $k$.
\begin{itemize}
\item[(R1)] {\it Continuity}: $D_1(A\vert \vert B)$ is a continuous function of $2n$ variables.
\item[(R2)] {\it Symmetry}: 
\begin{eqnarray*}
&&\hspace*{-18mm}  D_1\left(   a_1,\cdots ,a_j,\cdots ,a_k,\cdots ,a_n \vert \vert    b_1,\cdots ,b_j,\cdots ,b_k,\cdots ,b_n   \right) \\
&&\hspace*{-18mm}  =D_1\left(   a_1,\cdots ,a_k,\cdots ,a_j,\cdots ,a_n  \vert \vert    b_1,\cdots ,b_k,\cdots ,b_j,\cdots ,b_n   \right). 
\end{eqnarray*}
%\begin{eqnarray}
%&& D_1\left(   a_1,\cdots ,a_j,\cdots ,a_k,\cdots ,a_n \right. \nonumber \\
%&& \left.  \hspace*{8mm}  \vert \vert    b_1,\cdots ,b_j,\cdots ,b_k,\cdots ,b_n   \right) \nonumber \\
%&& = D_1\left(   a_1,\cdots ,a_k,\cdots ,a_j,\cdots ,a_n  \right. \nonumber \\ 
%&& \left. \hspace*{8mm}   \vert \vert    b_1,\cdots ,b_k,\cdots ,b_j,\cdots ,b_n   \right). 
%\end{eqnarray}
\item[(R3)] {\it Additivity}:
\begin{eqnarray*}
 &&  D_1 \left( a_{11} , \cdots ,a_{1m} ,a_{21} , \cdots ,a_{2m} \vert \vert  b_{11} , \cdots ,b_{1m} ,b_{21} , \cdots ,b_{2m}  \right)  = D_1 \left( {c_1 ,c_2 \left|\left|  {d_1 ,d_2 } \right.\right.    } \right) \\ 
 && + c_1 D_1 \left( {\frac{{a_{11} }}{{c_1 }}, \cdots ,\frac{{a_{1m} }}{{c_1 }} \left|\left|  {\frac{{b_{11} }}{{d_1 }}, \cdots ,\frac{{b_{1m} }}{{d_1 }}} \right.\right.  } \right)
+ c_2  D_1 \left( {\frac{{a_{21} }}{{c_2 }}, \cdots ,\frac{{a_{2m} }}{{c_2 }}\left|\left|  {\frac{{b_{21} }}{{d_2 }}, \cdots ,\frac{{b_{2m} }}{{d_2 }}} \right.\right.  } \right)  \label{Hob_H3}
 \end{eqnarray*}
where $c_i = \sum_{j=1}^m a_{ij}$ and $d_i = \sum_{j=1}^m b_{ij}$.
\item[(R4)]$D_1(A\vert \vert B) =0$ if $a_j =b_j$ for all $j$.
\item[(R5)]  $D_1(\frac{1}{n},\cdots ,\frac{1}{n},0,\cdots ,0\vert\vert  \frac{1}{n_0},\cdots ,\frac{1}{n_0})$ is an increasing function of $n_0$ and a decreasing function of $n$,
for any integers $n$ and $n_0$ such that $n_0 \geq n$.
\end{itemize}
\end{The}

As a one-parameter extension, we gave the uniqueness theorem for the Tsallis relative entropy as follows.
The function $D_q$ is defined for the probability distributions $A=\left\{a_j\right\}$ and $B=\left\{b_j\right\}$ on a finite probability space with one parameter $q \geq 0$. 
The one-parameter extended relative entropy (Tsallis relative entropy) was characterized by means of the following triplet of the generalized conditions (OR1), (OR2) and (OR3).  
\begin{Ax}{\bf (\cite{Furu1})} \label{g_hobson_axiom}

\begin{itemize}
\item[(OR1)] {\it Continuity}: $D_q(a_1,\cdots ,a_n \vert \vert  b_1, \cdots ,b_n)$ is a continuous function of $2n$ variables.
\item[(OR2)] {\it Symmetry}: 
\begin{eqnarray*}
&&\hspace*{-18mm}  D_q\left(   a_1,\cdots ,a_j,\cdots ,a_k,\cdots ,a_n \vert \vert    b_1,\cdots ,b_j,\cdots ,b_k,\cdots ,b_n   \right) \\
&&\hspace*{-18mm}  =D_q\left(   a_1,\cdots ,a_k,\cdots ,a_j,\cdots ,a_n  \vert \vert    b_1,\cdots ,b_k,\cdots ,b_j,\cdots ,b_n   \right). 
\end{eqnarray*}
%\begin{eqnarray*}
%&& D_q\left(   a_1,\cdots ,a_j,\cdots ,a_k,\cdots ,a_n   \right. \\
%&& \left. \hspace*{8mm}   \vert \vert    b_1, \cdots ,b_j,\cdots ,b_k,\cdots ,b_n   \right) \\
%&&= D_q\left(   a_1,\cdots ,a_k,\cdots ,a_j,\cdots ,a_n  \right. \\
%&& \left. \hspace*{8mm}   \vert \vert    b_1,  \cdots ,b_k,\cdots ,b_j,\cdots ,b_n   \right) 
%\end{eqnarray*}
\item[(OR3)] {\it Additivity}:
\begin{eqnarray}
&& D_q \left(   a_{11},\cdots ,a_{1m},\cdots ,a_{n1},\cdots ,a_{nm}  \vert \vert  b_{11},  \cdots ,b_{1m},\cdots ,b_{n1},\cdots ,b_{nm}   \right)   =  D_q(c_1,\cdots ,c_n \vert \vert  d_1 \cdots ,d_n)   \nonumber \\
&&      + \sum_{i=1}^n c_i^q d_i^{1-q} D_q\left( \frac{a_{i1}}{c_i},\dots ,\frac{a_{im}}{c_i} \left|\left|  \frac{b_{i1}}{d_i},\dots ,\frac{b_{im}}{d_i} \right.\right.  \right),\label{Tsa_H3}
\end{eqnarray} 
where $c_i = \sum_{j=1}^m a_{ij}$ and $d_i = \sum_{j=1}^m b_{ij}$.
\end{itemize}
\end{Ax}

Then, we have the following theorem.

\begin{The}\label{the-one-para} {\bf (\cite{Furu1})}
If conditions (OR1), (OR2) and (OR3) hold, then $D_q(A\vert B)$ is given in the following form:
\begin{equation}
D_q(A\vert\vert  B) = \sum_{j=1}^n  \frac{ a_j - a_j^q b_j^{1-q}}{\phi (q)}
\end{equation}
with a certain constant $\phi(q)$ depending on the parameter $q$. 
\end{The}

As for properties and applications of the Tsallis relative entropy, see our previous papers \cite{FYK,Furu2,Furu3}.

\section{Uniqueness theorem for two-parameter extended relative entropy} \label{sec3}
In our previous paper \cite{Furu1}, we gave Axiom \ref{g_hobson_axiom} in order to characterize the Tsallis relative entropy (one-parameter extended relative entropy).
In this section, we prove the uniqueness theorem for a two-parameter extended relative entropy.

\begin{The}\label{the1} 
If the function $D_{\alpha,\beta}(X\vert\vert  Y)$, defined for any pairs of the probability distributions $X=\left\{x_1,\cdots,x_n\right\}$ and $Y=\left\{y_1,\cdots,y_n\right\}$
on a finite probability space, satisfies the conditions (TR1)-(TR3) in the below, 
then  $D_{\alpha,\beta}(X \vert\vert  Y)$ is uniquely given by the form
\begin{equation}   \label{eq_the1}
D_{\alpha,\beta}(X\vert\vert  Y) = \sum_{j=1}^n  \frac{x_j^{\alpha}y_j^{1-\alpha} - x_j^{\beta} y_j^{1-\beta}}{\phi (\alpha,\beta)}
\end{equation}
with a certain constant $\phi(\alpha,\beta)$ depending on two parameters $\alpha$ and $\beta$.
\begin{itemize}
\item[(TR1)] {\it Continuity} : $D_{\alpha,\beta}  (x_1,\cdots ,x_n \vert \vert  y_1, \cdots ,y_n)$ is a continuous function for $2n$ variables.
\item[(TR2)] {\it Symmetry} : 
\begin{eqnarray*}
&&  D_{\alpha,\beta}(   x_1,\cdots ,x_j,\cdots ,x_k,\cdots ,x_n   \vert \vert   y_1,\cdots ,y_j,\cdots ,y_k,\cdots ,y_n   ) \\
&&    = D_{\alpha,\beta}(   x_1,\cdots ,x_k,\cdots ,x_j,\cdots ,x_n   \vert \vert  y_1,\cdots ,y_k,\cdots ,y_j,\cdots ,y_n   ). 
\end{eqnarray*}
\item[(TR3)] {\it Additivity} :
\begin{eqnarray}
&& D_{\alpha,\beta}(   x_{11},\cdots ,x_{1m},\cdots ,x_{n1},\cdots , x_{nm}   \vert \vert  y_{11},\cdots ,y_{1m},\cdots ,y_{n1},\cdots ,y_{nm}   ) \nonumber \\
&&  = D_{\alpha,\beta}(z_1,\cdots ,z_n \vert \vert  w_1 \cdots ,w_n) \sum_{j=1}^m \left(\frac{x_{ij}}{z_i}\right)^{\beta} \left(\frac{y_{ij}}{w_i}\right)^{1-\beta} \nonumber \\
&& + \sum_{i=1}^n z_i^{\alpha} w_i^{1-\alpha} D_{\alpha,\beta}\left( \frac{x_{i1}}{z_i},\dots ,\frac{x_{im}}{z_i} \left|\left|  \frac{y_{i1}}{w_i},\dots ,\frac{y_{im}}{w_i} \right.\right.  \right),\nonumber \\ \label{TR3}
\end{eqnarray} 
where $z_i = \sum_{j=1}^m x_{ij}$ and $w_i = \sum_{j=1}^m y_{ij}$.
\end{itemize}
\end{The}
{\it Proof}:
From (TR2), we have
\begin{eqnarray*}
 &&D_{\alpha ,\beta } \left( \frac{1}{{su}}, \cdots ,\frac{1}{{su}},0, \cdots ,0, \cdots ,\frac{1}{{su}}, \cdots ,\frac{1}{{su}},0, \cdots ,0 \left|\left|  \frac{1}{{tv}}, \cdots ,\frac{1}{{tv}} \right. \right.    \right) \\ 
 && = D_{\alpha ,\beta } \left( \frac{1}{{su}}, \cdots ,\frac{1}{{su}}, \cdots ,\frac{1}{{su}}, \cdots ,\frac{1}{{su}}, 0, \cdots ,0, \cdots ,0, \cdots ,0\left|\left|  \frac{1}{{tv}}, \cdots ,\frac{1}{{tv}} \right. \right.  \right). \\ 
 \end{eqnarray*}
From (TR3), we also have
\begin{eqnarray*}
&& D_{\alpha ,\beta } \left( \frac{1}{{su}}, \cdots ,\frac{1}{{su}},0, \cdots ,0, \cdots ,\frac{1}{{su}}, \cdots ,\frac{1}{{su}}, 0, \cdots ,0\left|\left|  \frac{1}{{tv}}, \cdots ,\frac{1}{{tv}} \right. \right.  \right) \\ 
&& \hspace*{-8mm}  = s\left( {\frac{1}{s}} \right)^\beta  \left( {\frac{1}{t}} \right)^{1 - \beta } D_{\alpha ,\beta } \left( {\frac{1}{u}, \cdots ,\frac{1}{u},0, \cdots ,0\left|\left|  {\frac{1}{v}, \cdots ,\frac{1}{v}} \right.\right.  } \right) \\
&& \hspace*{-8mm} + u\left( {\frac{1}{u}} \right)^\alpha  \left( {\frac{1}{v}} \right)^{1 - \alpha } D_{\alpha ,\beta } \left( {\frac{1}{s}, \cdots ,\frac{1}{s},0, \cdots ,0\left|\left|  {\frac{1}{t}, \cdots ,\frac{1}{t}} \right.\right.  } \right). \\ 
\end{eqnarray*}
From above two equations, we have
\begin{eqnarray*}
&&  D_{\alpha ,\beta } \left( \frac{1}{{su}}, \cdots ,\frac{1}{{su}}, \cdots ,\frac{1}{{su}}, \cdots ,\frac{1}{{su}}, 0, \cdots ,0, \cdots ,0, \cdots ,0\left|\left|  \frac{1}{{tv}}, \cdots ,\frac{1}{{tv}} \right. \right.  \right) \\ 
&&  = \left( {\frac{s}{t}} \right)^{1 - \beta } D_{\alpha ,\beta } \left( {\frac{1}{u}, \cdots ,\frac{1}{u},0, \cdots ,0\left|\left|  {\frac{1}{v}, \cdots ,\frac{1}{v}} \right.\right.  } \right) \\
&& + \left( {\frac{u}{v}} \right)^{1 - \alpha } D_{\alpha ,\beta } \left( {\frac{1}{s}, \cdots ,\frac{1}{s},0, \cdots ,0\left|\left|  {\frac{1}{t}, \cdots ,\frac{1}{t}} \right.\right.  } \right). \\ 
\end{eqnarray*}
If we put
\[
f_{\alpha ,\beta } \left( {s,t} \right) \equiv D_{\alpha ,\beta } \left( {\frac{1}{s}, \cdots ,\frac{1}{s},0, \cdots ,0 \left|\left|  {\frac{1}{t}, \cdots ,\frac{1}{t}} \right.  \right.} \right),\left( {t \ge s} \right),
\]
then we have
\[
f_{\alpha ,\beta } \left( {su,tv} \right) = \left( {\frac{s}{t}} \right)^{1 - \beta } f_{\alpha ,\beta } \left( {u,v} \right) + \left( {\frac{u}{v}} \right)^{1 - \alpha } f_{\alpha ,\beta } \left( {s,t} \right).
\]
We also have
\[
f_{\alpha ,\beta } \left( {us,vt} \right) = \left( {\frac{v}{u}} \right)^{1 - \beta } f_{\alpha ,\beta } \left( {s,t} \right) + \left( {\frac{s}{t}} \right)^{1 - \alpha } f_{\alpha ,\beta } \left( {u,v} \right),
\]
putting $s=u,u=s,t=v$ and $v=t$ in the above equation.
From above two equations, we have
\[
\frac{{\left( {\frac{s}{t}} \right)^{1 - \alpha }  - \left( {\frac{s}{t}} \right)^{1 - \beta } }}{{f_{\alpha ,\beta } \left( {s,t} \right)}} = \frac{{\left( {\frac{u}{v}} \right)^{1 - \alpha }  - \left( {\frac{u}{v}} \right)^{1 - \beta } }}{{f_{\alpha ,\beta } \left( {u,v} \right)}} \buildrel \Delta \over = \phi \left( {\alpha ,\beta } \right).
\]
Therefore we have
\[
f_{\alpha ,\beta } \left( {s,t} \right) = \frac{{\left( {\frac{s}{t}} \right)^{1 - \alpha }  - \left( {\frac{s}{t}} \right)^{1 - \beta } }}{{\phi \left( {\alpha ,\beta } \right)}}.
\]

For two natural numbers $l_i$ and $m_i$ such that $l_i \leq m_i$, we put
\[
z_i  \equiv \frac{{l_i }}{{\sum\limits_{k = 1}^n {l_k } }}, \left( {i = 1, \cdots ,n} \right),\quad 
w_i  \equiv \frac{{m_i }}{{\sum\limits_{k = 1}^n {m_k } }},   \left( {i = 1, \cdots ,n} \right)
\]
and
\begin{eqnarray*}
&&x_{ij}  \equiv \frac{1}{{\sum\limits_{k = 1}^n {l_k } }},\left( {i = 1, \cdots ,n;j = 1, \cdots ,l_i } \right),\\
&&y_{ij}  \equiv \frac{1}{{\sum\limits_{k = 1}^n {m_k } }},\left( {i = 1, \cdots ,n;j = 1, \cdots ,m_i } \right).
\end{eqnarray*}
From (TR2) and (TR3), we then have
\begin{eqnarray*}
&& D_{\alpha ,\beta } \left( \frac{1}{{\sum\limits_{k = 1}^n {l_k } }}, \cdots ,\frac{1}{{\sum\limits_{k = 1}^n {l_k } }},0, \cdots ,0, \cdots ,\frac{1}{{\sum\limits_{k = 1}^n {l_k } }}, \cdots ,\frac{1}{{\sum\limits_{k = 1}^n {l_k } }}, 0, \cdots ,0\left|\left|  \frac{1}{{\sum\limits_{k = 1}^n {m_k } }}, \cdots ,\frac{1}{{\sum\limits_{k = 1}^n {m_k } }} \right. \right. \right) \\ 
&& \hspace*{-8mm}   = D_{\alpha ,\beta } \left( {z_1 , \cdots ,z_n \left|\left|  {w_1 , \cdots ,w_n } \right.\right.} \right)\sum\limits_{j = 1}^{l_i } {\left( {\frac{1}{{l_i }}} \right)^\beta  \left( {\frac{1}{{m_i }}} \right)^{1 - \beta } } \\
&&\hspace*{-8mm}   + \sum\limits_{i = 1}^n {z_i^\alpha  w_i^{1 - \alpha } } D_{\alpha ,\beta } \left( {\frac{1}{{l_i }}, \cdots ,\frac{1}{{l_i }},0, \cdots ,0\left|\left|  {\frac{1}{{m_i }}, \cdots ,\frac{1}{{m_i }}} \right.\right. } \right), 
\end{eqnarray*}
since $x_{ij}=0$ for $j=l_i+1,\cdots ,m$.
Thus we have
\begin{eqnarray*}
&& D_{\alpha ,\beta } \left( {z_1 , \cdots ,z_n \left|\left|  {w_1 , \cdots ,w_n } \right.\right.} \right)  = \frac{{f_{\alpha ,\beta } \left( {\sum\limits_{k = 1}^n {l_k } ,\sum\limits_{k = 1}^n {m_k } } \right) - \sum\limits_{i = 1}^n {z_i^\alpha  w_i^{1 - \alpha } f_{\alpha ,\beta } \left( {l_i ,m_i } \right)} }}{{\sum\limits_{j = 1}^{l_i } {\left( {\frac{1}{{l_i }}} \right)^\beta  \left( {\frac{1}{{m_i }}} \right)^{1 - \beta } } }} \\ 
&& = \frac{  \left(\frac{  \sum_{k=1}^{n}l_k   }{\sum_{k=1}^{n}m_k   }\right)^{1-\alpha} -\left(\frac{  \sum_{k=1}^{n}l_k   }{\sum_{k=1}^{n}m_k   }\right)^{1-\beta}     }{\phi(\alpha,\beta)\left(\frac{l_i}{m_i}\right)^{1-\beta}} 
 - \frac{\sum_{i=1}^n z_i^{\alpha} w_i^{1-\alpha} \left\{ \left(\frac{l_i}{m_i}\right)^{1-\alpha} -\left(\frac{l_i}{m_i}\right)^{1-\beta}  \right\}}{\phi(\alpha,\beta)\left(\frac{l_i}{m_i}\right)^{1-\beta}} 
 \end{eqnarray*}
Here we have 
$$
\sum\limits_{i = 1}^n {z_i^r w_i^{1 - r} \left( {\frac{{l_i }}{{m_i }}} \right)^{1 - r} }   = \sum\limits_{i = 1}^n {\left( {\frac{{l_i }}{{\sum\limits_{k = 1}^n {l_k } }}} \right)^r \left( {\frac{{m_i }}{{\sum\limits_{k = 1}^n {m_k } }}} \right)^{1 - r} \left( {\frac{{l_i }}{{m_i }}} \right)^{1 - r} }  = \left( {\frac{{\sum\limits_{k = 1}^n {l_k } }}{{\sum\limits_{k = 1}^n {m_k } }}} \right)^{1 - r} ,\left( {r \in \mathbb{R}} \right)
$$
for 
\[
z_i  \equiv \frac{{l_i }}{{\sum\limits_{k = 1}^n {l_k } }},  \left( {i = 1, \cdots ,n} \right),\quad
w_i  \equiv \frac{{m_i }}{{\sum\limits_{k = 1}^n {m_k } }},  \left( {i = 1, \cdots ,n} \right).
\]
Thus we have
$$
D_{\alpha ,\beta } \left( {z_1 , \cdots ,z_n \left|\left|  {w_1 , \cdots ,w_n } \right.\right.    } \right) 
= \frac{{\sum\limits_{i = 1}^n {z_i^\alpha  w_i^{1 - \alpha } } \left( {\frac{{l_i }}{{m_i }}} \right)^{1 - \beta }  - \sum\limits_{i = 1}^n {z_i^\beta  w_i^{1 - \beta } \left( {\frac{{l_i }}{{m_i }}} \right)^{1 - \beta } } }}{{\phi \left( {\alpha ,\beta } \right)\left( {\frac{{l_i }}{{m_i }}} \right)^{1 - \beta } }}.
$$
Since we can take $l_i$ and $m_i$ arbitrary, we may take $l_i=l$ and $m_i=m$, then we have
\[
D_{\alpha ,\beta } \left( {z_1 , \cdots ,z_n \left|\left|  {w_1 , \cdots ,w_n } \right.\right.  } \right) 
= \frac{{\sum\limits_{i = 1}^n {z_i^\alpha  w_i^{1 - \alpha } }  - \sum\limits_{i = 1}^n {z_i^\beta  w_i^{1 - \beta } } }}{{\phi \left( {\alpha ,\beta } \right)}}.
\]
From (TR1) and the fact that any real number can be approximated by a rational number, the above result is true for any positive real number $z_j$ and $w_j$
satisfying $\sum_{j=1}^n z _j = \sum_{j=1}^n w_j =1$. 

\hfill \qed

Putting $\beta = 1$ and $\alpha=q$ in the above theorem, we have the uniqueness theorem for a one-parameter extended relative entropy (Theorem \ref{the-one-para}).

%%%%%%%%%%%%%%%%%%%%%%%%%%%%%%%%%%%%%%%%%%%%%%%%%%%%%%%%%%%%%%%%
%%%%%%%%%%%%%%%%%%%%%%%%%%%%%%%%%%%%%%%%%%%%%%%%%%%%%%%%%%%%%%%
%%%%%%%%%%%%%%%%%%%%%%%%%%%%%%%%%%%%%%%%%%%%%%%%%%%%%%%%%%%%%%%
%%%%%%%%%%%%%%%%%%%%%%%%%%%%%%%%%%%%%%%%%%%%%%%%%%%%%%%%%%%%%%%%
%%%%%%%%%%%%%%%%%%%%%%%%%%%%%%%%%%%%%%%%%%%%%%%%%%%%%%%%%%%%%%%
%%%%%%%%%%%%%%%%%%%%%%%%%%%%%%%%%%%%%%%%%%%%%%%%%%%%%%%%%%%%%%%

\section{Characterizations of $\phi(\alpha,\beta)$}
In this section, we characterize the constant $\phi(\alpha,\beta)$ depending on two parameters $\alpha$ and $\beta$.

\begin{Prop}
The postulate that our quantity $D_{\alpha ,\beta } \left( {x_1 , \cdots ,x_n \left|\left|  {y_1 , \cdots ,y_n } \right. \right.} \right)$ 
defined for any pairs of the probability distributions: 
\begin{equation} \label{postulate01}
D_{\alpha ,\beta } \left( {x_1 , \cdots ,x_n \left|\left|  {y_1 , \cdots ,y_n} \right.\right.} \right)
 = \sum\limits_{j = 1}^n {\frac{{x_j^\alpha  y_j^{1 - \alpha }  - x_j^\beta  y_j^{1 - \beta } }}{{\phi \left( {\alpha ,\beta } \right)}}} 
\end{equation}
derived in Theorem \ref{the1} recovers the relative entropy
 when $\alpha \to 1$ and $\beta \to 1$, that is,
\begin{equation} \label{postulate02}
\mathop {\lim }\limits_{\alpha ,\beta  \to 1} D_{\alpha ,\beta } \left( {x_1 , \cdots ,x_n \left|\left|  {y_1 , \cdots ,y_n} \right.\right.} \right)
 = k\sum\limits_{j = 1}^n {x_j \left(\log x_j -\log y_j\right)} 
\end{equation}
implies the following conditions.
\begin{itemize}
\item[(c1)] We have $
\mathop {\lim }\limits_{\alpha  \to 1} \phi \left( {\alpha ,1} \right) = \mathop {\lim }\limits_{\beta  \to 1} \phi \left( {1,\beta } \right) 
= \mathop {\lim }\limits_{\beta  \to \alpha } \phi \left( {\alpha ,\beta } \right) = 0$
and $\phi \left( {\alpha ,\beta } \right) \ne 0\,\,for\,\,\alpha  \ne \beta$.
\item[(c2)] There exists the interval $(a,b)$ such that $\phi(\alpha,1)$  and  $\phi(1,\beta)$ are differentiable on  $\left( {a,1} \right) \cup \left( {1,b} \right)$.
\item[(c3)]There exists the constant $k>0$  such that  $
\mathop {\lim }\limits_{\alpha  \to 1} \frac{{d\phi \left( {\alpha ,1} \right)}}{{d\alpha }} = \frac{1}{k}$
and
$
\mathop {\lim }\limits_{\beta  \to 1} \frac{{d\phi \left( {1,\beta } \right)}}{{d\beta }} =  - \frac{1}{k}$.
\end{itemize}
\end{Prop}

{\it Proof}:
\begin{itemize}
\item[(c1)] 
We may calculate the limit of the left hand side in Eq.(\ref{postulate02}) in the following ways.
\begin{itemize}
\item[(i)]
Firstly we may take the limit $\alpha\to 1$ in Eq.(\ref{postulate01}) 
and then later take the limit $\beta \to 1$:
$$
\hspace*{-15mm} \mathop {\lim }\limits_{\beta  \to 1} D_{1,\beta } \left( {x_1 , \cdots ,x_n \left|\left|  {y_1 , \cdots ,y_n} \right.\right.} \right)
 = \mathop {\lim }\limits_{\beta  \to 1} \sum\limits_{j = 1}^n {\frac{{x_j  - x_j^\beta  y_j^{1 - \beta } }}{{\phi \left( {1,\beta } \right)}}}. 
$$
Since we have
$\mathop {\lim }\limits_{\beta  \to 1} \sum\limits_{j = 1}^n {\left( {x_j  - x_j^\beta  y_j^{1 - \beta } } \right)}  = 0$, 
we need $\mathop {\lim }\limits_{\beta  \to 1} \phi \left( {1,\beta } \right) = 0$ in order that we have the limit in the above.
\item[(ii)] By the similar way to (i), we have $\mathop {\lim }\limits_{\alpha  \to 1} \phi \left( {\alpha ,1} \right) = 0$.
\item[(iii)] Firstly we may put $\beta \to \alpha$ 
and then later take the limit $\alpha\to 1$.
In the case $\beta \to \alpha$, the summation of the numerator of the right hand side in Eq.(\ref{postulate01}) is equal to $0$:
$$\lim_{\beta \to \alpha}\sum_{j=1}^n \left(x_j^{\alpha} y_j^{1-\alpha}-x_j^{\beta} y_j^{1-\beta}\right)=0.$$ 
Therefore we have $\lim_{\beta\to\alpha}\phi(\alpha,\beta)=0$, otherwise 
$\lim_{\beta\to\alpha}D_{\alpha,\beta}(x_1,\cdots,x_n\vert\vert  y_1,\cdots,y_n)$ takes $0$, which contradicts the Eq.(\ref{postulate02}).
From the reason why we have the limit of the left hand side in (\ref{postulate02}),
 we also have $\phi \left( {\alpha ,\beta } \right) \ne 0$ for $ \alpha \neq \beta$, since $
\sum\limits_{j = 1}^n {x_j^\alpha  y_j^{1 - \alpha }  - x_j^\beta  y_j^{1 - \beta } }  \ne 0$ for $ \alpha \neq \beta$.
\end{itemize}
\item[(c2)] Since $
\sum\limits_{j = 1}^n {x_j  - x_j^\beta  y_j^{1 - \beta } } $ is differentiable by $\beta$, we need that there exists an interval $(a,b)$ such that
$\phi(1,\beta)$ is also differentiable by $\beta$ on $\left( {a,1} \right) \cup \left( {1,b} \right)$, in order that we have the limit of the left hand side in Eq.(\ref{postulate02}).
By the similar way,  there exists an interval $(a,b)$ such that
$\phi(\alpha,1)$ is also differentiable by $\beta$ on $\left( {a,1} \right) \cup \left( {1,b} \right)$.
\item[(c3)] Since we have
\begin{eqnarray*}
\mathop {\lim }\limits_{\beta  \to 1} D_{1,\beta } \left( {x_1 , \cdots ,x_n \left|\left|  {y_1 , \cdots ,y_n} \right.\right.  } \right) 
&=& \mathop {\lim }\limits_{\beta  \to 1} \sum\limits_{j = 1}^n {\frac{{x_j  - x_j^\beta  y_j^{1 - \beta } }}{{\phi \left( {1,\beta } \right)}}}\\
&=& \mathop {\lim }\limits_{\beta  \to 1} \frac{{ - \sum\limits_{j = 1}^n {x_j^\beta  y_j^{1 - \beta } \left( {\log x_j  - \log y_j } \right)} }}{{\frac{{d\phi \left( {1,\beta } \right)}}{{d\beta }}}},
\end{eqnarray*} 
there exists a constant $k>0$ such that $
\frac{{d\phi \left( {1,\beta } \right)}}{{d\beta }} =  - \frac{1}{k}$.
By the similar way, 
there exists a constant $k>0$ such that $
\frac{{d\phi \left( {\alpha,1 } \right)}}{{d\beta }} =   \frac{1}{k}$.
\end{itemize}

\hfill \qed

\begin{Prop}
$D_{\alpha,\beta}(X \vert\vert  U)$ takes the minimum value for fixed posterior probability distribution as uniform distribution 
$U=\left\{\frac{1}{n},\cdots ,\frac{1}{n}  \right\}$ :
$$D_{\alpha,\beta}\left(x_1,\cdots ,x_n \left|\left|  \frac{1}{n},\cdots ,\frac{1}{n}\right.\right.  \right) 
\geq D_{\alpha,\beta}\left(\frac{1}{n},\cdots ,\frac{1}{n}  \left|\left|  \frac{1}{n},\cdots ,\frac{1}{n}\right.\right.  \right), $$
when we have  
\begin{itemize}
\item[(c4)] the following relations  (i) and (ii) for $\alpha$ and $\beta$
\begin{itemize}
\item[(i)] $\alpha\neq\beta$.
\item[(ii)]
  If $\phi \left( {\alpha ,\beta } \right) > 0$, then we have $0 \le \beta  \le 1 \le \alpha$.  
  If $\phi \left( {\alpha ,\beta } \right) < 0$, then we have $0 \le \alpha  \le 1 \le \beta$. 
\end{itemize}

\end{itemize}

\end{Prop}
{\it Proof}:
The second derivative of $
D_{\alpha ,\beta } \left( {x_1 , \cdots ,x_n \left|\left|  {\frac{1}{n}, \cdots ,\frac{1}{n}} \right.\right.} \right)
$ on $x_j$ is calculated by 
$$
 \frac{{d^2 D_{\alpha ,\beta } \left( {x_1 , \cdots ,x_n \left|\left|  {\frac{1}{n}, \cdots ,\frac{1}{n}} \right.\right.} \right)}}{{dx_j^2 }}  = \frac{{n^{\alpha  - 1} \alpha \left( {\alpha  - 1} \right)x_j^{\alpha  - 2}  - n^{\beta  - 1} \beta \left( {\beta  - 1} \right)x_j^{\beta  - 2} }}{{\phi \left( {\alpha ,\beta } \right)}}
$$
This takes positive value in the case of (c4) so that it should be convex in $x_j$. Therefore $D_{\alpha,\beta}(X\vert\vert  U)$ takes the minimum value. 

\hfill \qed

%%%%%%%%%%%%%%%%%%%%%%%%%%%%%%%%%%%%%%%%%%%%%%%%%%%%%%%%%%%%%%%%%%%%%%%%%%%%%%%%
%%%%%%%%%%%%%%%%%%%%%%%%%%%%%%%%%%%%%%%%%%%%%%%%%%%%%%%%%%%%%%%%%%%%%%%%%%%%%%%%
%%%%%%%%%%%%%%%%%%%%%%%%%%%%%%%%%%%%%%%%%%%%%%%%%%%%%%%%%%%%%%%%%%%%%%%%%%%%%%%%
%%%%%%%%%%%%%%%%%%%%%%%%%%%%%%%%%%%%%%%%%%%%%%%%%%%%%%%%%%%%%%%%%%%%%%%%%%%%%%%%
%%%%%%%%%%%%%%%%%%%%%%%%%%%%%%%%%%%%%%%%%%%%%%%%%%%%%%%%%%%%%%%%%%%%%%%%%%%%%%%%
%%%%%%%%%%%%%%%%%%%%%%%%%%%%%%%%%%%%%%%%%%%%%%%%%%%%%%%%%%%%%%%%%%%%%%%%%%%%%%%%
%%%%%%%%%%%%%%%%%%%%%%%%%%%%%%%%%%%%%%%%%%%%%%%%%%%%%%%%%%%%%%%%%%%%%%%%%%%%%%%%
%%%%%%%%%%%%%%%%%%%%%%%%%%%%%%%%%%%%%%%%%%%%%%%%%%%%%%%%%%%%%%%%%%%%%%%%%%%%%%%%

\section{Properties of a two-parameter extended relative entropy}
As an example satisfying  the conditions (c1)-(c4) on $\phi(\alpha,\beta)$, we simply take $\phi(\alpha,\beta) = \alpha -\beta$.
Then we may define a two-parameter extended relative entropy in the following.
\begin{Def}
For two parameters $\alpha,\beta\in  \mathbb{R}$ satisfying $ 0 \le \alpha  \le 1 \le \beta $
or  $0 \le \beta  \le 1 \le \alpha$, and 
two probability distributions $X=\left\{x_1,\cdots,x_n\right\}$ and $Y=\left\{y_1,\cdots,y_n\right\}$,
we define a two-parameter extended relative entropy by
\[
D_{\alpha ,\beta } \left( {X \left| \left|  {Y} \right.\right.} \right) 
\equiv \sum\limits_{j = 1}^n {\frac{{x_j^\alpha  y_j^{1 - \alpha }  - x_j^\beta  y_j^{1 - \beta } }}{{\alpha  - \beta }}},\quad (\alpha\neq\beta). 
\]
\end{Def}
Note that a two-parameter extended relative entropy is a generalization of the relative entropy in the sense that
$$
\lim_{\alpha,\beta\to 1}D_{\alpha ,\beta } \left( {X \left|\left|  {Y} \right.\right.} \right) = D_1(X\vert \vert Y).
$$
We also note that a two-parameter extended relative entropy recovers 
the Tsallis relative entropy (one-parameter extended relative entropy)
when $\alpha=1$ or $\beta=1$. 
The Tsallis relative entropy is also a one-parameter generalization of the relative entropy:
$$\lim_{q\to 1}D_{q}^T(X \vert \vert  Y) = D_1(X\vert \vert Y).$$

In addition, we note that a two-parameter extended relative entropy  is expressed by the convex combination of the Tsallis relative entropy:
\begin{equation}\label{relation}
D_{\alpha,\beta}(X \vert \vert  Y) = \frac{\alpha-1}{\alpha-\beta}D_{\alpha}^T(X \vert \vert  Y)+\frac{1-\beta}{\alpha-\beta}D_{\beta}^T(X \vert \vert  Y).
\end{equation}

Thus we have the following properties on a two-parameter relative entropy, thanks to the above relation and the properties of the Tsallis relative entropy studied in \cite{FYK}.
\begin{Prop}\label{prop_1}
For a two-parameter extended relative entropy $D_{\alpha,\beta}(X \vert \vert  Y)$, we have the following properties.
\begin{itemize}
\item[(i)] (Nonnegativity) $D_{\alpha,\beta}  (X \vert \vert  Y) \geq 0$.
\item[(ii)] (Symmetry) 
$$
 D_{\alpha,\beta} \left( {x_{\pi \left( 1 \right)} , \cdots ,x_{\pi \left( n \right)} \left|\left|  {y_{\pi \left( 1 \right)} , \cdots ,y_{\pi \left( n \right)} } \right.\right.} \right) 
 = D_{\alpha,\beta} \left( {x_1 , \cdots ,x_n \left|\left|  {y_1 , \cdots ,y_n} \right.\right.} \right).
$$
\item[(iii)] (Possibility of extension) 
$$
 D_{\alpha,\beta} \left( {x_1 , \cdots ,x_n ,0 \left|\left|  {y_1 , \cdots ,y_n,0} \right.\right.} \right) 
 = D_{\alpha,\beta} \left( {x_1 , \cdots ,x_n \left|\left|  {y_1 , \cdots ,y_n} \right.\right.} \right).
$$
\item[(iv)] (Joint convexity) For $0 \leq \lambda \leq 1$ and the probability distributions $X^{(i)}=\left\{x_j^{(i)}\right\}$,$Y^{(i)}=\left\{y_j^{(i)}\right\}$, $(i=1,2;j=1,\cdots,n)$,  we have 
\begin{eqnarray*}
\hspace*{-10mm}  &&D_{\alpha,\beta} \left( \lambda X^{\left( 1 \right)}  + \left( {1 - \lambda } \right)X^{\left( 2 \right)} \left|\left|  \lambda Y^{\left( 1 \right)}  + \left( {1 - \lambda } \right)Y^{\left( 2 \right)}  \right.\right.\right) \\
\hspace*{-10mm} &&\le \lambda D_{\alpha,\beta} \left( {X^{\left( 1 \right)} \left|\left|  Y^{\left( 1 \right)} \right.\right. } \right) + \left( {1 - \lambda } \right)D_{\alpha,\beta} \left( {X^{\left( 2 \right)} \left|\left|  Y^{\left( 2 \right)} \right.\right.} \right).
\end{eqnarray*}
\item[(v)] (Monotonicity)  For the transition probability matrix $W$, we have
\[D_{\alpha,\beta} \left( {WX \left| \left| {WY} \right.\right.} \right) \le D_{\alpha,\beta} \left( {X \left|\left|  Y \right.\right.} \right).\]
\end{itemize}
\end{Prop}

It is also notable that we have the following expression for a two-parameter extended relative entropy:
$$
D_{\alpha,\beta}(X\vert\vert U) = \frac{n^{\alpha-1}-n^{\beta -1}}{\alpha-\beta} -n^{\alpha -1} \left( \frac{\alpha -1}{\alpha-\beta}\right) S_{\alpha}(X) -
n^{\beta -1} \left(\frac{1-\beta}{\alpha-\beta} \right)S_{\beta}(X) 
$$
for the uniform distribution $U=\left\{1/n,\cdots,1/n\right\}$,
while we also have the following relation between the two-parameter extended entropy and the Tsallis entropy (one-parameter extended entropy):
$$
S_{\alpha,\beta}(X) = \left( \frac{\alpha -1}{\alpha-\beta}\right) S_{\alpha}(X) + \left( \frac{1-\beta}{\alpha-\beta} \right) S_{\beta}(X). 
$$
Therefore we may not obtain the direct relation between $D_{\alpha,\beta}(X\vert\vert U)$ and $S_{\alpha,\beta}(X)$ except for $n=1$.
%However, if we consider a simple distribution such as $I=\left\{ 1,\cdots,0\right\}$ for example, we easily find their relation as
%$$S_{\alpha,\beta}(X) = - D_{\alpha,\beta}(X\vert\vert I). $$

\section{Conclusion} 

As we have seen in Section \ref{sec3}, the two-parameter extended relative entropy is characterized by continuity, symmetry and additivity.
On the other hand, it is known that the $f$-divergence is characterized by symmetry, monotonicity and joint convexity \cite{Csi2}. 
The properties such as monotonicity and joint convexity are represented by the inequalities.
For the characterization of $f$-divergence, we need the inequalities involving their equality conditions, while for the characterization
of  the two-parameter extended relative entropy, we need the functional equation referred by an additivity.  
Therefore the conditions in our axiom are essentially different from those of the axiom characterizing $f$-divergence. 
It is also notable that our characterization of a two-parameter extended relative entropy (Theorem 3.1) is the uniqueness theorem
such that the function $D_{\alpha,\beta}(X\vert\vert Y)$ is uniquely given by Eq.(\ref{eq_the1}), while
the characterization of $f$-divergence (Theorem 1 in \cite{Csi2}) is the existence theorem for a convex function $f$ such that
the function defined for any pair of the probability distributions is equal to the $f$-divergence.
In other words, in the paper \cite{Csi2}, the  existence of the convex function has been shown but
 the uniqueness of the convex function $f$ has not been shown, 
so that our axiomatic characterization may have an advantage since it uniquely gives a two-parameter extended relative entropy.
It is also notable that the uniqueness theorem for $\alpha$-divergence was recently shown in \cite{Ama} for the special case such that
the divergence measure (functional) is written by a sum of all components.

Closing this section, we give the expressions of a two-parameter extended relative entropy by means of $f$-divergence:
$$
D_f(X\vert\vert Y)\equiv \sum_{j=1}^n y_jf\left(\frac{x_j}{y_j}\right),
$$
where $f$ is a convex function on $(0,\infty)$ and $f(1)=0$.
If we take $f(t)=t \log t$, then $f$-divergence $D_f(X\vert\vert Y)$ recovers the relative entropy.
Here, if we put 
\begin{equation} \label{f-div_func01}
f_{\alpha,\beta}(t) \equiv \frac{t^{\alpha}-t^{\beta}}{\alpha-\beta}, \quad (\alpha \neq \beta),
\end{equation}
then
$\frac{d^2f_{\alpha,\beta}(t)}{dt^2} \geq 0$ for $0 \le \alpha  \le 1 \le \beta $
or  $0 \le \beta  \le 1 \le \alpha$.
And then we have the following expression:
$$
D_{\alpha,\beta}(X\vert\vert Y) = D_{f_{\alpha,\beta}}(X\vert\vert Y).
$$

It is known that the relative entropy is connected to many important results in the mathematical physics and information science.
For a two-parameter extended relative entropy, such connections (for example with H-theorem or variational expressions related to the free energy) will be studied in the future.

%The $f$-divergence is often defined by
%$$D_{f^*}(X\vert\vert Y)\equiv \sum_{j=1}^n x_jf^*\left(\frac{y_j}{x_j}\right),$$
%where $f$ is a convex function on $(0,\infty)$ and $f(1)=0$.
%If we take $f^*(t)=- \log t$, then $f$-divergence $D_{f^*}(X\vert\vert Y)$ recovers the relative entropy.
%Here, if we put 
%\begin{equation} \label{f-div_func02}
%f^{*}_{\alpha,\beta}(t) \equiv \frac{t^{1-\alpha}-t^{1-\beta}}{\alpha-\beta}, \quad (\alpha \neq \beta),
%\end{equation}
%then $\frac{d^2f^{*}_{\alpha,\beta}(t)}{dt^2} \geq 0$ for $0 \le \alpha  \le 1 \le \beta $or  $0 \le \beta  \le 1 \le \alpha$.
%Indeed, if the function  $f(t)$ is convex on $(0,\infty)$, then the function $g(t)\equiv t f\left(\frac{1}{t}\right)$ is also convex on $(0,\infty)$, because of the elementary calculation:
%$$\frac{d^2g(t)}{dt^2} = \frac{1}{t^3}\frac{d^2f\left(\frac{1}{t}\right)}{dt^2}.$$
%And then we have the following expression:
%$$D_{\alpha,\beta}(X\vert\vert Y) = D_{f^{*}_{\alpha,\beta}}(X\vert\vert Y).$$
%It is known that the dual of $f$-divergence:
%$$D^*_f(X\vert\vert Y) \equiv D_f(Y\vert \vert X)$$
%has the relation 
%$$D^*_f(X\vert\vert Y) = D_{f^*}(X\vert \vert Y),$$
%if we have $f^*(t)=tf(\frac{1}{t})$. (For example, see \cite{AN} for details.)
%In the case of two-parameter extended relative entropy, the above relation holds:
%$f^*_{\alpha,\beta}(t) = t f_{\alpha,\beta} (\frac{1}{t})  $.
%Through the concept of duality in the field of information geometry, we find the relation between Eq.(\ref{f-div_func01}) and Eq.(\ref{f-div_func02}).

\section*{Acknowledgements}
I  would like to thank Professor H.Suyari and Professor T.Wada giving me an opportunity to read their interesting paper \cite{WS} in the workshop at Chiba University.
The author was partially supported by the Japanese Ministry of Education, Science, Sports and Culture, 
Grant-in-Aid for Encouragement of Young Scientists (B) 20740067.

\end{document}